\documentclass[9pt,twocolumn,twoside]{osajnl}

\journal{ol} 

\setboolean{shortarticle}{true}

\title{Linear detection of 30 mW dual-comb interferograms}

\author[1]{Philippe Guay}
\author[1]{Mathieu Walsh}
\author[1]{Alex Tourigny-Plante}
\author[1,*]{Jérôme Genest}

\affil[1]{Centre d'optique, photonique et laser, Universit\'{e} Laval, Qu\'{e}bec, Qu\'{e}bec G1V 0A6, Canada}

\affil[*]{Corresponding author: jerome.genest@copl.ulaval.ca}

\begin{abstract}
Detector nonlinearity is an important factor limiting the maximal power and hence the signal-to-noise ratio (SNR) in dual-comb interferometry. To increase the SNR without overwhelming averaging time, specific experimental conditions must be met to ensure that photodetector nonlinearity is properly handled for high input power. Detectors exhibiting nonlinear behavior can produce linear dual-comb interferograms if the area of the detector's impulse response does not saturate and if the overlap between successive time-varying impulse responses is properly managed. Here, a high bandwidth non-amplified photodetector is characterized in terms of its impulse response to high intensity short pulses to exemplify the conditions. With 30 mW of continuous power on the detector, nonlinear spectral artifacts in dual-comb interferograms are at least 35 dB below the signal. A comparative spectroscopic measurement with a frequency swept laser shows that no systematic transmittance error can be attributed to nonlinearity. 
\end{abstract}

\setboolean{displaycopyright}{true}

\begin{document}

\maketitle

Amplified photodetectors such as the PDB series from Thorlabs exhibit nonlinear behavior with short-pulses when their output has reached saturation. As incident power is increased, the amplitude of the detector's impulse response reaches a plateau and significantly broadens \cite{GUA21a}. In dual-comb experiments \cite{COD16}, this behavior is observed as the detection of nonlinear artifacts in the interferogram's spectrum \cite{GUA21b}. Such artifacts distort the signal's spectrum in which absorption lines can be measured. As a result, incorrect lines intensities are retrieved. To prevent such nonlinear systematic errors, a correction algorithm has been suggested \cite{GUA21c}. This procedure is based on the existing classical Fourier transform spectrometer methods that minimize out-of-band spectral artifacts \cite{LAC00}. It has been shown that this nonlinearity correction enables short-time measurements with high SNR without overwhelming integration time \cite{GUA21c,ROY12}. 

As an alternative to post-processing nonlinearity correction, experimental conditions can be wisely chosen to handle photodetector nonlinearity. As such, the detector's bandwidth has to allow the detector's impulse responses to be separated. This way, any nonlinear change in the shape of the detector's response can be taken into account. Moreover, the use of a detector without built-in amplification removes a prominent nonlinearity source and reduces measurement noise, provided that sufficient optical power is available to produce an output voltage measurable with high dynamic range using commercial acquisition instruments. 

In this letter, a non-amplified balanced detector is characterized in terms of its nonlinear behavior. The impulse responses of both photodiodes are measured for various incident power levels. It is shown that the area of the impulse responses has a linear relation with the input power up to 35 mW, making the retrieval of linear dual-comb interferograms possible even if the shape of the impulse response changes as a result of nonlinearity.  Dual-comb measurements are presented to expose the absence of nonlinear spectral artifacts up to 35 dB below the signal level.  An HITRAN comparison of  H$^{12}$CN spectral transmittance measurements acquired both with the dual-comb interferometer and a swept tunable laser validates the linearity of the results. Residuals transmittance errors are confirmed to be systematic spectroscopic model errors. 

\begin{figure}[htbp]
\centering
\includegraphics[width=0.5\textwidth]{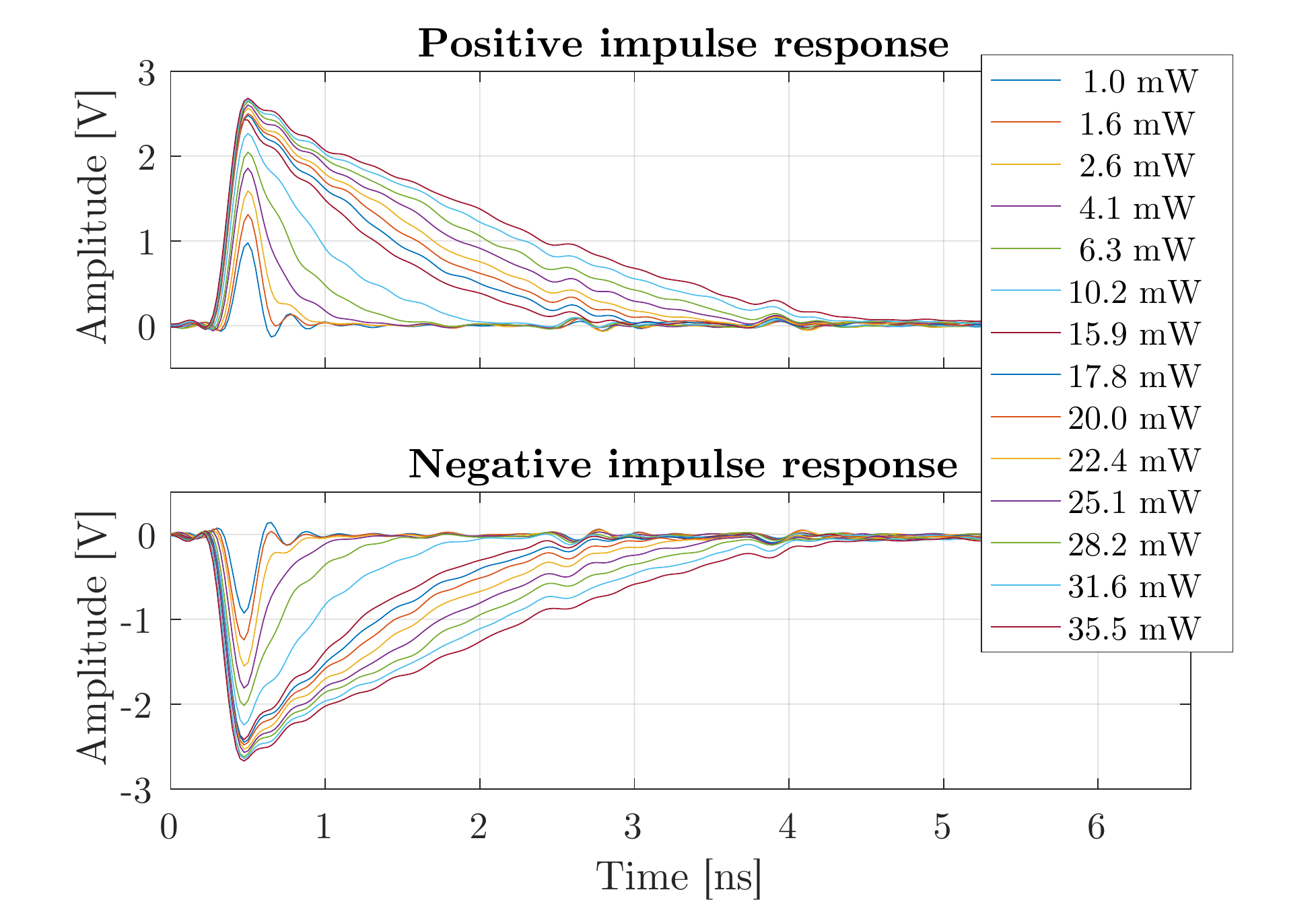}
\caption{\label{fig:imp_resp} Impulse responses for the positive and the negative photodiodes of Thorlabs’ balanced detector BDX1BA up to 35 mW.}
\end{figure}

The impulse response of the detector's positive and negative photodiodes (Thorlabs BDX1BA) for various incident optical powers are shown in Fig. \ref{fig:imp_resp}. The pulses were sent to a variable optical attenuator before reaching the detector. Impulse responses were digitized for varying optical power up to 35 mW, which is 25~mW above the maximum input power specified by the manufacturer. The impulse responses were acquired with a 10~GHz bandwidth oscilloscope. 

The shape of the impulse responses is broadening with increased power as it has been previously demonstrated in \cite{DID09}. This effect is explained by the excess of photocarriers which causes a screening of the electric field that slows down the carriers and reduces the photodetector's bandwidth \cite{WIL94}. Even if the shape of the impulse response displays a nonlinear behavior with increasing power, the area of the impulse response contains in fact the information of interest for each point of the dual-comb interferogram. 

It is important to understand the fact that the detector's impulse responses are varying with power and that it can introduce a dynamic nonlinearity if successive pulses are overlapped. The value of an interferogram point will be affected by the amplitude of the preceding pulse. Therefore, nonlinearity can be handled in a dual-comb measurement by first making sure that the photodetector's impulse responses do not overlap such that the area of each pulse can be individually retrieved. A properly designed linear and stationary low pass filter keeping only the first spectral alias is then used to compute the area in each repetition period (here $f_r =160$ MHz, $T_r=6.25$ ns). In practice, the width of the non-stationary photodetector impulse responses must always be appreciably smaller than the duration of the stationary filter to properly mitigate the dynamic nonlinearity. 

\begin{figure}[ht!]
\centering
\includegraphics[width=0.5\textwidth]{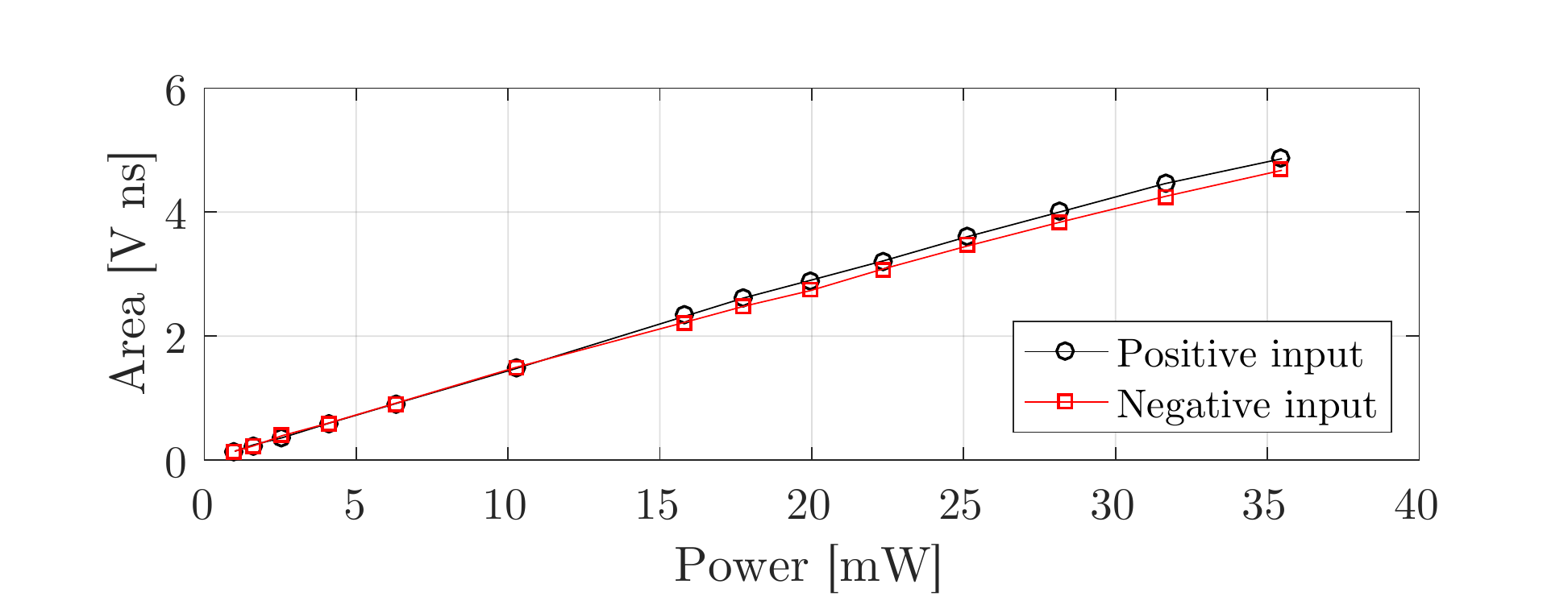}
\caption{\label{fig:area} Area of the impulse response for both photodiodes of the BDX1BA detector as a function of the averaged incident optical power.}
\end{figure}

The area of the impulse response is computed and plotted against the incident power in Fig. \ref{fig:area}. For both photodiodes, the relation appears somewhat linear, thus hinting that the detector's nonlinearity will minimally affect the measurement as long as the area of each pulse is properly obtained. The different slopes in the figure indicates that the matching of photodiodes is imperfect, but that can be  corrected by adjusting the power sent to each photodiode in a balanced measurement. In cases where the optical power to impulse response area relations are insufficiently linear, a static nonlinearity correction method can be used \cite{GUA21c}.

\begin{figure}[htbp]
\centering
\includegraphics[width=0.5\textwidth]{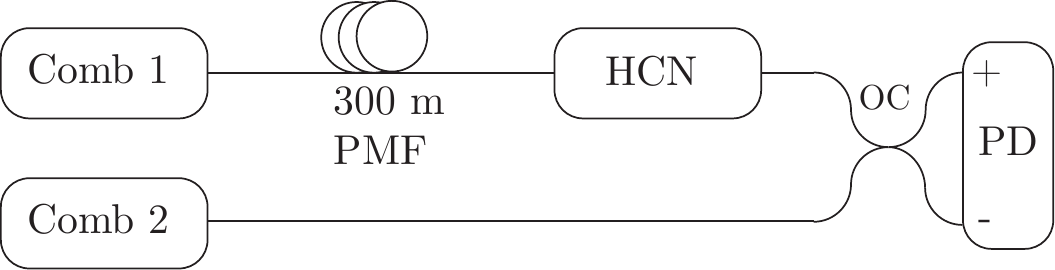}
\caption{\label{fig:setupDCS} Block diagram of the dual-comb setup. PMF: Polarization-maintaining fiber, HCN: H$^{12}$CN gas cell. OC: 50/50 Output coupler, PD : Balanced photodetector}
\end{figure}

Dual-comb interferograms were acquired using two passively mode-lock lasers at 1550~nm based on semiconductor saturable absorber mirrors \cite{SIN15}. The experimental setup is shown in Fig. \ref{fig:setupDCS}. A dispersive fiber was placed in the setup to chirp the pulses and thus avoid saturation of the acquisition card, thereby reducing nonlinear effects. It has also been specifically placed in the gas cell arm to avoid any Fano asymmetry in the absorption lines caused by delta-like excitation of gas molecules \cite{GUA22b}. A H$^{12}$CN gas cell was used to measure absorption lines listed in the HITRAN database. 

\begin{figure}[htbp]
\centering
\includegraphics[width=0.5\textwidth]{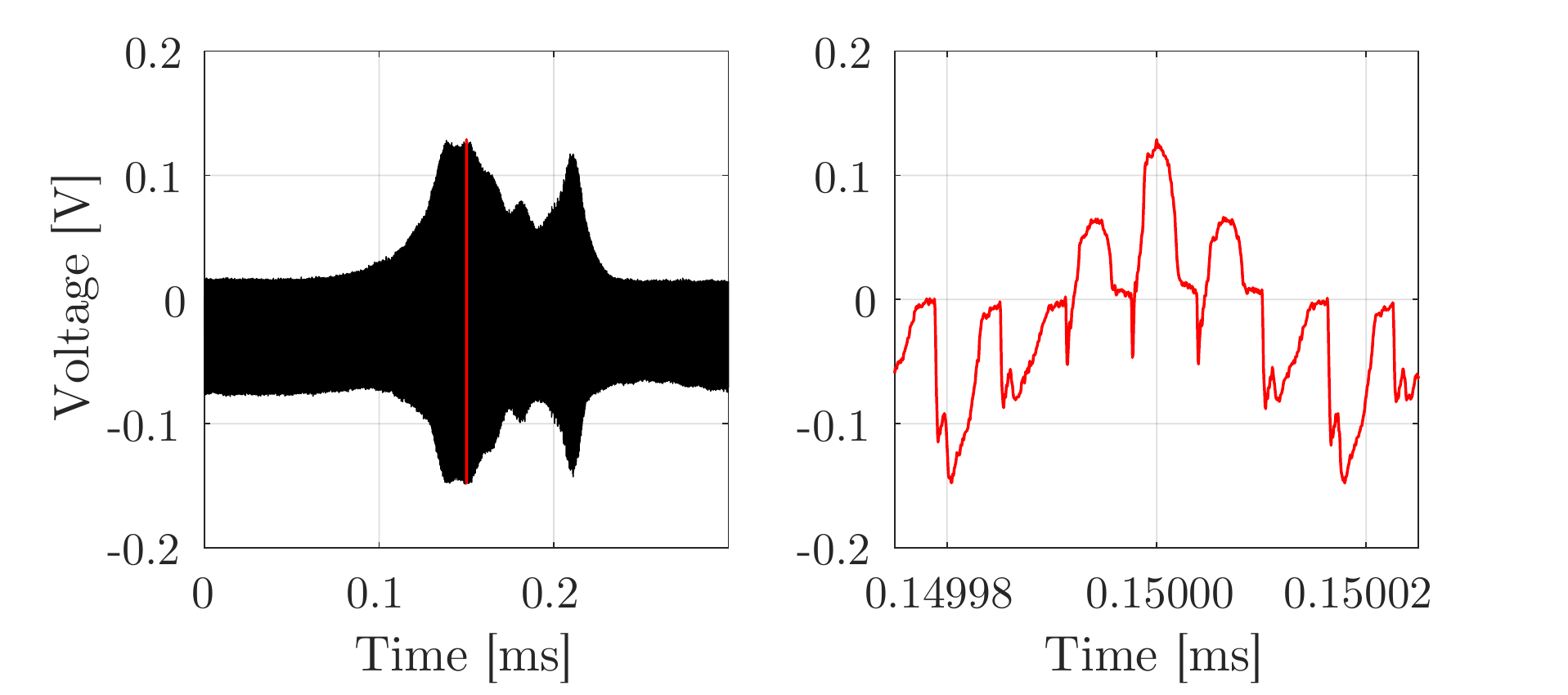}
\caption{\label{fig:ZPD_pulses} (left) Interferogram's centerburst at ZPD. (right) Close-in on the separated impulse response in the ZPD.}
\end{figure}

A chirped interferogram centerburst is shown on the left of Fig.\ref{fig:ZPD_pulses}, while the right side panel shows a close-in on a few pulses within the burst. This illustrates the impulse responses separation for 30 mW incident power. The interferogram centerburst shows that this measurement is slightly unbalanced as negative pulses appear dominant outside the zero path diffrence (ZPD) burst, meaning that the negative photodiode receives stronger pulses that the positive one. This is confirmed by computing the low-pass (80 MHz) filtered interferogram which has a non-zero mean.  This has no impact on interferogram linearity as these additional unbalanced terms lead only to extra DC component to the filtered interferogram. It is worth mentioning that the interferogram measured voltage at 30 mW is much lower than the characterized pulses voltage around 2.5 V since interferograms are significantly chirped. 

In this measurement, the impulse responses appear broadened, but the laser repetition rate is sufficiently low to maintain the independence of the pulses. The impulse responses shown in Fig.\ref{fig:ZPD_pulses} appear to have a different shape than in the characterization in Fig. \ref{fig:imp_resp}. This is explained by an imperfect interferogram modulation efficiency that leaves pulses on both photodiodes even in a ZPD fringe where power should be directed at the positive photodiode in a positive ZPD fringe for instance. Since pulses are detected on both photodiodes, the measured pulses are the result of the subtraction of impulse responses and thus present a slightly different shape, but remains separated.

To quantify the presence of nonlinearity, one can look for the nonlinear spectral artifacts. Since a nonlinear interferogram ($\text{IGM}_{\text{NL}}$) can be written as the series expansion of the linear interferogram ($\text{IGM}_{\text{L}}$) as 

\begin{equation}\label{poly}
\text{IGM}_{\text{NL}} =  a_0 + a_1[\text{IGM}_{\text{L}}]  + a_2[\text{IGM}_{\text{L}}]^2  + ...
\end{equation}

where the $a_i$ are the nonlinear coefficients, the interferogram's spectrum is expected to have a contribution resulting from the self-convolution of the linear spectrum. For instance, the 2nd nonlinear term will introduce a spectral contribution at DC and at twice the frequency of the signal. Similarly, the 3rd order spectral artifact will have a contribution at the frequency of interest and at three times the frequency. A complete description of the nonlinear model is given in \cite{GUA21b}. 

In a first measurement, a low 140 Hz repetition rate difference ($\Delta f_r$) was set to offer a clear view of the spectral artifacts. The interferogram's spectrum was wisely placed around 15 MHz by tuning the frequency locks with the optical reference to allow a clear visualisation of the 2nd and 3rd order artifacts at 30 and 45 MHz respectively. Only the central portion containing the zero-path-difference burst is processed. 

The impact of nonlinearity is exacerbated in the interferogram centerburst where the signal explores the largest dynamic range. To quantify small amounts of nonlinearity against noise, it is therefore advantageous to compute a short Fourier transform around the centerburst before looking for spectral content at the 2nd and 3rd harmonics of the signal. Here, the 300 m chirping fiber produces temporally spread interferograms. In an effort to maximize the signal-to-noise ratio of this nonlinearity assessment, the interferogram is first unchirped, thus concentrating all the nonlinearity impact in a few points around zero path difference and a Fourier-transform on only 1000 points sampled at 125 MHz is computed, which is actually shorter than the chirped centerburst. This may seems counter intuitive in the time domain, but this processing does not change the spectral content, apart for its phase and a better SNR.

\begin{figure}[htbp]
\centering
\includegraphics[width=0.5\textwidth]{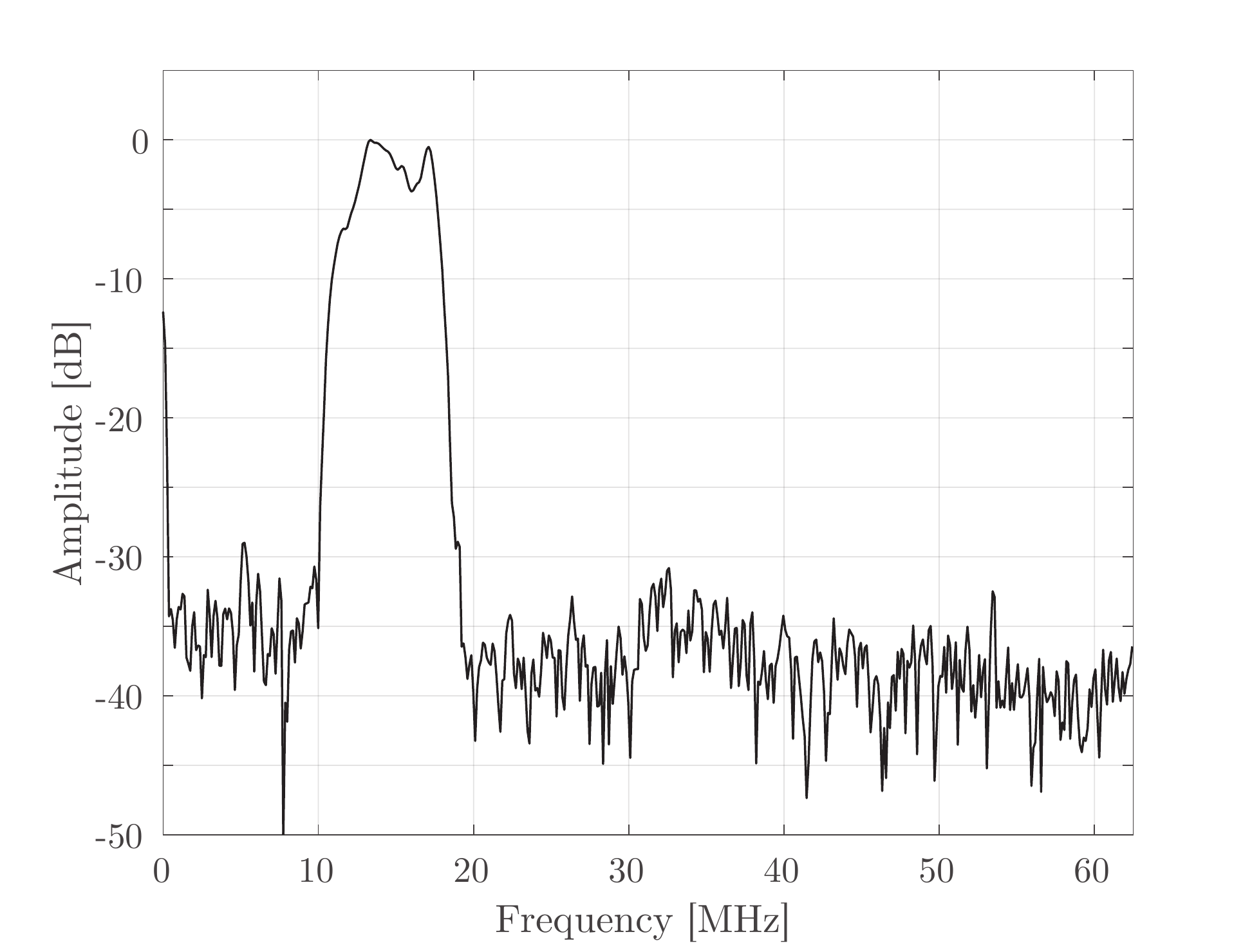}
\caption{\label{fig:DCS_lowdfr} Low resolution dual-comb spectrum showing a wake mode at 34 MHz 30 dB below the signal level and no spectral artifacts at 30 and 45 MHz at 35 dB below the signal level.}
\end{figure}

The spectrum of an interferogram's central portion is shown in Fig. \ref{fig:DCS_lowdfr}. It can readily be seen that spectral artifacts are not visible at a 30 dB level below the signal of interest. The hump centered at 34 MHz could be interpreted as second order nonlinearity that would somehow be slightly shifted from 30 MHz, but this is not the case. This is actually a contribution from the laser wake mode \cite{WAN17,ROZ19}. This has been confirmed by shifting the interferogram's spectrum and showing that the hump is always at a constant frequency offset from the signal rather than being at an harmonic of the signal as nonlinear spectral artifacts would. This means that nonlinear artifacts are actually below the 35 dB level. 

A second round of measurements with an higher repetition rate difference was performed. Using a higher $\Delta f_r$ of about 300 Hz, the signal of interest is spread across the $f_r/2$ band and the nonlinear artifacts, if any, are aliased and folded on the signal of interest. If nonlinearity is significant, it should thus appear as systematic errors on the calibrated transmittance.   

The measured spectrum was fitted to a model using a sum of Voigt profiles. An optimization procedure is performed to retrieve the cell parameters (pressure, length and temperature) as well as the optical point spacing and the absolute frequency of the dataset. The inteferograms have been digitized with a Gage acquisition card that has a 125 MS/s sampling rate and thus required a low-pass filter at 62.5 MHz. This filter is the one used to keep only one spectral alias and thus computes the area of each pulse. Ten datasets of 400 ms have been digitized, phase-corrected using a 1562 nm reference laser (RIO Planex) and phase-corrected with a self-correction algorithm \cite{HEB17} to remove out-of-loop residual phase noise. The datasets have then been aligned in phase and delay by a cross-correlation before averaging to yield a 4 s total measurement time. 

\begin{figure*}[htbp]
\centering
\includegraphics[width=\textwidth]{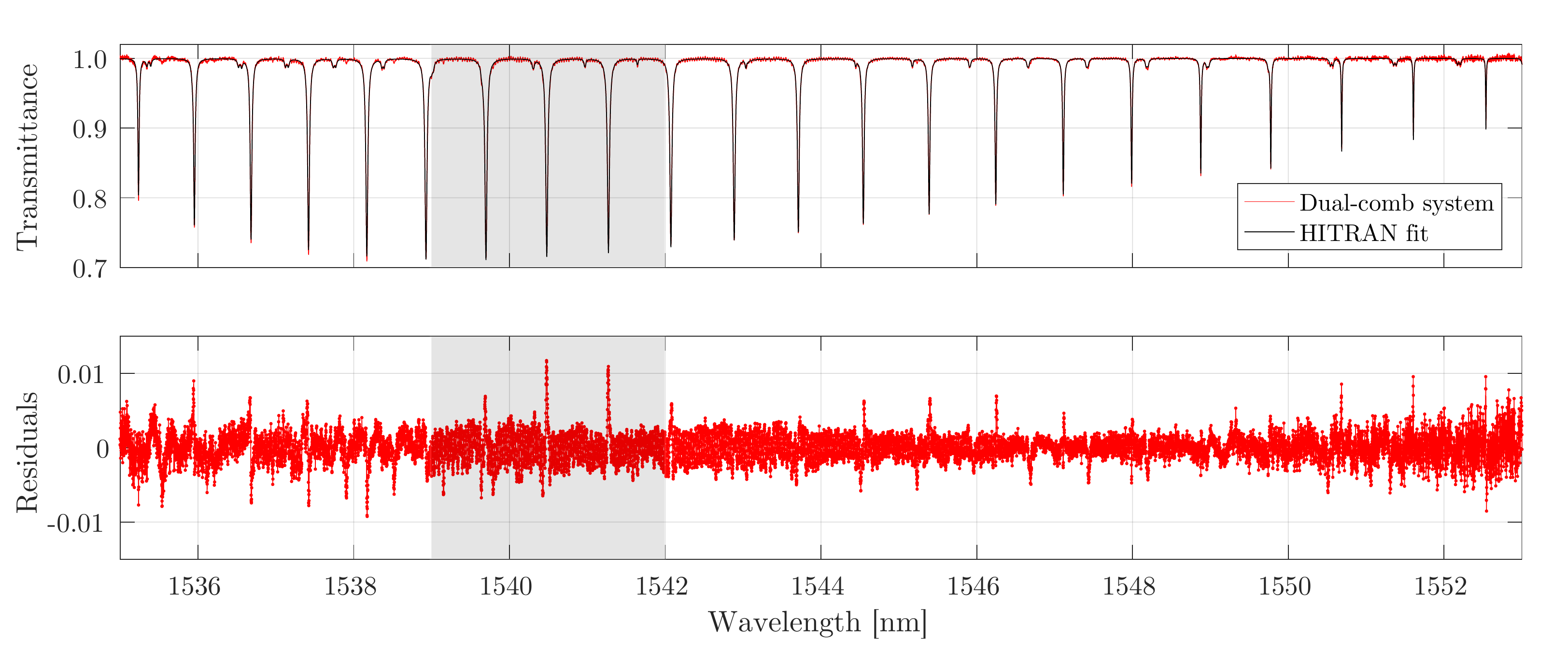}
\caption{\label{fig:spectroDCS}(top panel) Transmittance spectrum of H$^{12}$CN (P branch) for a 30 mW measurement and as modeled with Voigt line shapes computed with parameters from HITRAN 2016 database (black). (bottom panel) Residuals between the experimental data and the theoretical modeling.}
\end{figure*}

The transmittance spectrum for a 30 mW measurement is shown in Fig. \ref{fig:spectroDCS}. The transmittance curves for the experimental data are shown in the top panel while the bottom panel shows the residuals between the HITRAN model and the data. 

Since the largest residuals in \ref{fig:spectroDCS} appear to be $\approx 1\%$ systematic errors on several of the spectral lines, the gas cell spectral transmittance was also measured using a tunable laser system (LUNA OVA-5000). This measurement, providing excellent SNR due to high power single-mode laser and due to the 100 averaged scans, is also compared to HITRAN, using a similar optimisation procedure and the same cell parameters.

\begin{figure}[htbp]
\centering
\includegraphics[width=0.5\textwidth]{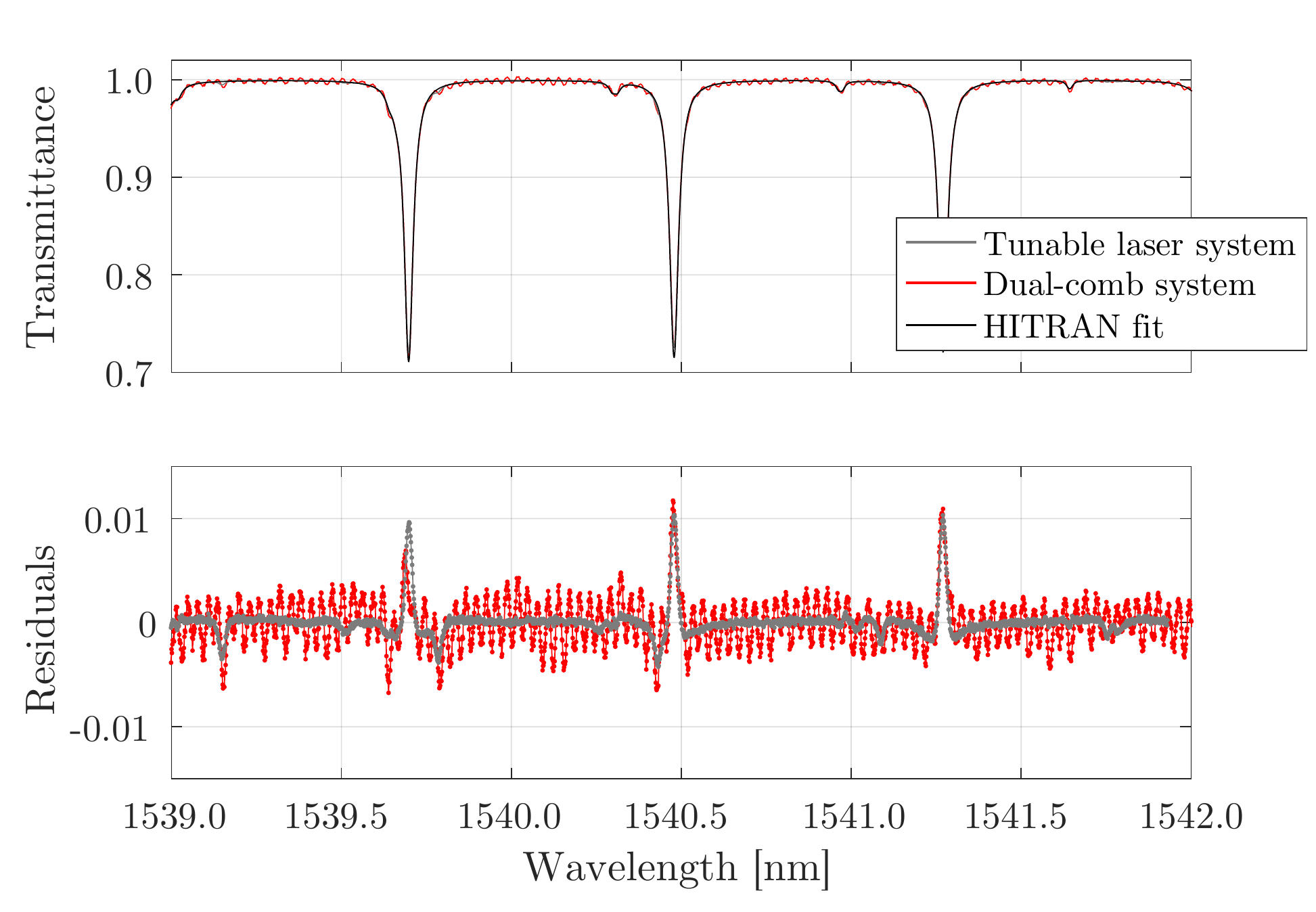}
\caption{\label{fig:resLUNA}(top panel) Transmittance spectrum for the dual-comb system (30 mW CW power), the tunable laser system and the HITRAN model. (bottom panel) Residuals between measurements.}
\end{figure}

Fig. \ref{fig:resLUNA} shows that the two experimentally measured transmittances have the same line centered systematic residuals when compared to HITRAN. This provides a convincing argument that these deviations are not caused by nonlinearity in the dual-comb measurement but instead arise from an imperfect spectroscopic model, pointing to spectral lineshapes beyond the Voigt profile \cite{TEN14}.  

The residuals of the dual-comb measurement also contain an etalon that is only observed when the optical pulses are chirped. Whether the pulses are chirped with hundred of meters of polarization-maintaining fiber or with a custom-made fiber Bragg grating, a similar etalon appears in both cases. This is attributed to differential delays and cross talk between polarisation axes in the chirping elements. At a level below $1\%$, this is compatible with the expected polarisation cross talk in the fiber specified at 25 dB/100 m. Chirping interferograms to alleviate dynamic range issues at detection thus comes with a price in the form of a systematic error and an optimized measurement needs to balance those factors. Here, this etalon is slightly larger than the random noise contribution but smaller than the dominant residuals arising from the non-Voigt profiles.


As a conclusion, linear dual-comb interferograms can be obtained from photodetectors operated in a nonlinear regime. To that effect, one has to ensure the power-dependence of the detector impulse response does not affect the estimation of each pulse area. This usually implies that the photodetector has a quoted bandwidth larger than $f_r$.  In the experimental demonstration, $30$ mW of continuous power is sent to an unamplified balanced detector pair, yielding interferograms with nonlinear 2nd and 3rd harmonics under 35 dB below the signal level, enabling measuring absorption lines of H$^{12}$CN with precision that could further the line modeling theory.


\begin{backmatter}
\bmsection{Funding} This work was supported by Natural Sciences and Engineering Research Council of Canada (NSERC), Fonds de Recherche du Québec - Nature et Technologies (FRQNT) and the Office of Sponsored Research (OSR) at King Abdullah University of Science and Technology (KAUST) via the Competitive Research Grant (CRG) program with grant \# OSR-CRG2019-4046

\bmsection{Acknowledgments} 

\bmsection{Disclosures} The authors declare no conflicts of interest.

\bmsection{Data availability} Data underlying the results presented in this paper are not publicly available at this time but may be obtained from the authors upon reasonable request.

\end{backmatter}

%




\end{document}